\documentclass[preprint,preprintnumbers]{revtex4}

\usepackage{graphicx}

\usepackage{amssymb}

\usepackage{mathrsfs}

\begin{document}

\noindent

\title{Thermal conductance  of   an  Edge mode }

\author{D. Schmeltzer}

\affiliation{Physics Department, City College of the City University of New York,  New York,New York 10031 USA}

\begin{abstract}

Thermoelectric conductance of an edge mode 
 is investigated .The  edge modes of  a   $2D $ and $3D$ two band  model with  parabolic dispersion is considered   .

\noindent

\noindent
 For the  the one dimensional  non interacting   fermions the   thermal conductivity computed  agrees with the result  known  from    $Bosonization$ computations. 

\noindent
 In  the presence of a magnetic field,  backscattering is allowed and      controls the  value of the  thermal conductivity .

\noindent
The thermal conductivity  is  obtained from the continuity equation  of thermal  current energy conserevation.  The thermal conductivity is  computed  introducing  the $Scattering$ matrix  for particles and anti-particles . 

\noindent
   At finite temperatures the   backscattering is allowed,  the  electric conductance ,the thermoelectric  conductance  and the   thermal conductance  decrease with the increase of the magnetic field. At finite temperatures,  weak localization effects  are  small   and can be ignored . We confirm the experimental results in a magnetic field for a $3D$ Topological Insulator. An  experimental set-up  was proposed to test our theory.

\end{abstract}

\maketitle

\textbf{I. INTRODUCTION}

Thermal conductance  is the flow of heat that results  from a   temperature gradient\cite{Goldsmith,Mahan,chiatti,Luttinger,Sivan,kane}. Thermoelectrics are used in cooler  refrigerators based on the Peltier effect  \cite{Goldsmith},which predicts the appearance of a heat current when an electric current passes through a material.  Alternatively, the Seebeck effect  generates an electric  current from a temperature gradient \cite{Kamran}. According to \cite{Mahan,Amnon}, the presence of a disorder  can  enhance the figure of merit \cite{Kamran}. These results have been obtained within  the Boltzmann theory. Recent experiments performed by  \cite{Pepper} suggest  that interference effects  are important and may invalidate the   Boltzmann theory.

\noindent
Topological aspects of low-energy excitations are revealed through  the  excitation of the  edge modes. Recently the thermal  conductance has been measured  for fractional Quantized Hall state \cite{Banerjee} which was investigated using the Bosonization  method  \cite{kane}. Here we will show that the thermal conductivity is obtained using a  direct Fermionic formulation.

\noindent
The goal   of this paper  is to  investigate the thermal conductance for an edge mode. As an  example we consider a two band  model with parabolic dispersion \cite{Zhou,Xiao,Shun} .  We consider  open boundary condition.
In  the $x$  direction   the boundary  are   at  $x=0$  , $x= 
D>1000nm$ and periodic in the $y$ direction.  For   finite widths $D>1000nm$  in the $x$ direction, it was shown that the  coupling for the different edges can be ignored \cite{Zhou}. As a result , we can replace  the two dimensional model   with a semi  infinite  
plane.  The    edge modes  for different  spins move  in opposite directions.The edge modes are  protected against backscattering and the thermal conductance  is given  by $\kappa=\nu\frac{\pi^2K^{2}_{B}T}{3h}$ with $\nu=1$.

The    paper is outlined as follows : in chapter II, we will present  the  two band model with parabolic dispersion. This model has   an edge mode on the   boundaries  at $x=0$ and $x=D>1000nm$ . 
 In chapter  III, 
we formulate  the   Landauer- Buttiker method    \cite{Buttiker,Flensberg,Butcher}  for  Dirac materials with particles and anti-particles. We    compute the  electric conductance $ G$, thermoelectric conductance $L$, and the  thermal conductance $\kappa$.  In chapter IV, we include a magnetic field and find that  backscattering is allowed. We compare the thermoelectric conductance with and without backscattering . We show that, as a  result of backscattering,  the electrical conductance,  thermoelectric  conductance and thermal conductance decrease with the magnetic field.  We also comment on  the electrical conductance  for a  $3D $  topological insulator in a magnetic field    with a   $2D$ boundary  surface \cite{Ando,Kozlov} at  finite temperatures  weak anti-localization effects can be ignored   \cite{Raman}.As a result bacscattering controls the conductance.
  In chapter V, we propose   an experimental set-up   for testing  this theory. In section VII we present our conclusion.

\vspace{0.2 in}

\textbf{II.  The edge  mode for a two band model with parabolic dispersion }

\vspace{0.2 in}

 We consider  a  two dimensional two band  model with parabolic dispersion   \cite{Xiao,Zhou,Shun}, given by:
\begin{eqnarray}
&&H=H_{0}+H_{1}\nonumber\\&&
H_{0}=\hbar v \int_{0}^{\infty}dx\int _{-\infty}^{\infty}dy \Psi^{\dagger}\Big[\tau_{3}\otimes\sigma_{1}(-i\partial_{1})+\tau_{1}\otimes I(\Delta-\beta\partial^{2}_{x})\Big]\Psi(x,y)\nonumber\\&&
H_{1}=\hbar v \int_{0}^{\infty}dx\int _{-\infty}^{\infty}dy \Psi^{\dagger}\Big[\tau_{3}\otimes \sigma_{2} (-i\partial_{2}) -I\otimes I k_{F}\Big]\Psi(x,y)\nonumber\\&&
\end{eqnarray}
where $\vec{\tau}$  is a Pauli which describes the two band model, $\beta^{-1}$ is proportional to the  effective mass.
We find the zero modes of the Hamiltonian $H_{0}$ and use the zero modes to project out the Hamiltonian $H_{1}$. The Hamiltonian   $H_{0}$  contains the non-relativistic (parabolic ) dispersion given by the term $(\Delta-\beta\partial^{2}_{x})$.
$ v$ is the  Fermi velocity ,$k_{F}$ is  the Fermi momentum and $\sigma_{2}$ is the Pauli matrix .
The spinor $\Psi(x,y)$ is given in terms of the two component  chiral fields  $ \Psi(x,y)=\Big[\Psi_{R}(x,y),\Psi_{L}(x,y)\Big]$.  We  use the Pauli matrix $\tau_{3}$  for the fields to replace the  spinor $\Psi(x,y) $ with the spinor $\hat{\Psi}(x,y)$. We  define: $ \Psi(x,y)=\tau_{3}\hat{\Psi}(x,y)$.

\noindent

The zero mode equation of the Hamiltonian   $H_{0}$  is given by:
\begin{eqnarray}
&&\Big[\sigma_{1}(-i\partial_{1})-i\tau_{2}(\Delta-\beta\partial^{2}_{x})\Big]\hat{\Psi}(x,y)=0\nonumber\\&&\
\end {eqnarray}
In Eq.$(2)$ we substitute, $\hat{\Psi}(x,y)=\eta_{\pm}(x) $ where  $\eta_{\pm}(x)=e^{\gamma x}\hat{\eta}_{\pm}$ .The    spinors solution for the edge modes obey : $\tau_{2}\sigma_{1}\hat{\eta}_{+}=\hat{\eta}_{+}$ and $\tau_{2}\sigma_{1}\hat{\eta}_{-}=-\hat{\eta}_{-}$.   $\gamma$ is  the root of the  quadratic equation  $ \gamma+\Delta -\beta\gamma^2=0$. 
The zero mode vanishes at $x=0$  and at $x\approx D>1000nm $( at  this limit the edges are decoupled) . For a particular choice of parameters $\beta<0$ ,$\Delta>0$ and $ \Delta \cdot  \beta>-1$ we find the spinors   for the edge modes.
\begin{eqnarray}
&&\eta_{+}(x)=\hat{\eta}_{+}\sqrt{\frac{1}{N}}\Big(e^{-\gamma_{1}x} -e^{-\gamma_{2}x}\Big),\hspace {0.1 in} \eta_{-}(x)=\hat{\eta}_{-}\sqrt{\frac{1}{N}}\Big(e^{-\gamma_{1}x} -e^{-\gamma_{2}x}\Big);\gamma_{1}(B,\Delta)>0,\gamma_{2}(B,\Delta)>0\nonumber\\&&,
\hat{\eta}_{+}=\frac{1}{2}\Big[1 ,1,i,i\Big]^{T} ,\hspace{0.05 in}\hat{\eta}_{-}=\frac{1}{2}\Big[1,-1,i,-i\Big]^{T} \nonumber\\&&
\end{eqnarray}
The space 
 $ 0\leq x= D>1000nm$,  where $x$  is approximated by  $ 0\leq x \rightarrow \infty$.
 $\sqrt{\frac{1}{N}}$ is the normalization factor which depends on the two roots $\gamma_{1}$ ,$\gamma_{2}$ and length $x=D$, for the limit $x\approx D>1000nm$ we  use $x\rightarrow \infty$ and find that the dependence on $\gamma_{1}$, $\gamma_{2}$ disappears  after the integration with respect $x$ which cancels the contribution from the normalization    factor $\sqrt{\frac{1}{N}}$. 
Next, we replaced the field $\Psi(x,y)$ with the projected spinor on     the  zero modes:
\begin{equation}
 \Psi(x,y)\approx \hat{\Psi}_{+}(y)\eta_{+}(x)+ \hat{\Psi}_{-}(y)\eta_{-}(x)
\label{projection}
\end{equation}
Using this projected basis, we obtain the explicit form of the $2D$ system of the projected $H_{1}$ Hamiltonian. 
\begin{eqnarray}
&&H= \hbar v\int_{0}^{D>1000nm}\,dx\int_{-\infty}^{\infty}dy\cdot\nonumber\\&&\Big[\hat{\Psi}^{\dagger}_{+}(y)\langle \eta_{+}(x)| (I \otimes\sigma_{2})|\eta_{+}(x) \rangle (-i\partial_{2})\hat{\Psi}_{+}(y) +\hat{\Psi}^{\dagger}_{-}(y)\langle \eta_{-}(x)|(I\otimes \sigma_{2})|\eta_{+}(x) \rangle (-i\partial_{2})\hat{\Psi}{+}(y)\nonumber\\&&+ \hat{\Psi}^{\dagger}_{+}(y)\langle \eta_{+}(x)| (I\times I(-k_{F} ) |\eta_{+}(x)\rangle\hat{\Psi}{+}(y)+\hat{\Psi}^{\dagger}_{-}(y)\langle \eta_{-}(x)| (I\otimes I(-k_{F} )| \eta_{-}(x)\rangle \hat{\Psi}{-}(y)\Big]\nonumber\\&&
\end{eqnarray}
Performing the integration with respect to the   $x$  coordinate we obtain  the $1D$ projected Hamiltonian. (The    explicit dependence on the $\gamma_{1}(\beta,\Delta)$ ,$\gamma_{2}(\beta,\Delta)$ cancel due to the normalization factor $\sqrt{\frac{1}{N}}$ when the $x$ integration is extended  to the interval $\Big[ 0 ,\infty\Big]$ .) We observe that due to the projection, the two dimensional model given in Eq.$(1)$ has  become a one dimensional  model given by Eq.$(6)$. We introduce  the notation  $C(y)=\Big[\hat{\Psi}_{+}(y),\hat{\Psi}_{-}(y)\Big]^{T}$ and find:
\begin{equation}
H=H_{0} +H_{1}\Rightarrow H^{edge}= \hbar v \int_{-\infty}^{\infty}dy \Big[C^{\dagger}(y)\Big( \sigma_{2}(-i\partial_{2}) -I k_{F}\Big)C(y)\Big]
\label{1d}
\end{equation}
Eq.$(6)$ represents an effective topological model in one dimensions which describes  a  two dimensional model with edges modes.

\vspace{0.2 in}

\textbf{III-The conserved electric and thermal current for the one dimensional topological model  obtained from the projected model  }

\vspace{0.2in} 
In this section we derive the electric and thermal currents using the continuity equation for the charge and energy density for the model given in Eq.$(6)$. This discussion is included since the formulation presented is not well known. In the second part of this section we consider the one dimensional model in the eigenvalue representation. This formulation  is needed for the use of the Landauer-Buttiker formalism.

\noindent
The electric  charge operator is given by $Q(y)=eC^{\dagger}(y)IC(y )$.  Using the continuity equation   $\partial_{t}Q(y,t)+\partial_{y}J^{el}(y,t)$ with  the  Heisenberg equation of motion, we find the electric current operator:
\begin{equation}
J^{el}(y,t)=ev C^{\dagger}(y,t)\sigma_{2}C(y ,t)
\label {electric}
\end{equation}

\noindent

 The thermal  energy  density relative to the Fermi energy is:
\begin{equation}
h(y,t)= \hbar v  \Big[C^{\dagger}(y)\Big( \sigma_{2}(-i\partial_{2}) -I k_{F}\Big)C(y)\Big]  \label{thermakl}
\end{equation}
\noindent
The heat density  obeys  the heat continuity equation,
$\partial_{t}h(y,t)+\partial_{y}J^{heat}(y,t)=0$

\noindent

From the Heisenberg equation of motion, we obtain the thermal heat current density $J^{heat}(y,t)$:
\begin{equation}
J^{heat}(y,t)=\hbar v^2 C^{\dagger}(y)\Big(-i\partial_{2} -\sigma_{2}k_{F}\Big)C(y )
\label {heat}
\end{equation}

\noindent
In  order to compute the currents, we need to determine the momentum expansion of the field operator $C(y ) $. For this purpose, we compute the eigenvalues and the eigen vectors for the energy density   $h(y,t)$. We find $\lambda^{+}=\hbar v |q|-E_{F}$
hich corresponds to the  eigen vector $ |U^{+}(q)\rangle =\sqrt{\frac{1}{2}}\Big[i(\theta[-q]-\theta[q]),1\Big]^{T}$ and the negative eigenvalue  $\lambda^{-}=-\hbar v |q|-E_{F}$ which corresponds to the eigen vector $ |V^{-}(q)\rangle =\sqrt{\frac{1}{2}}\Big[-i(\theta[-q]-\theta[q]),1\Big]^{T}$.  $\theta[q]$ represents the step function which takes values of  $1$ or $0$ .The momentum expansion for the field operator is given in terms of the  particle operators $a^{\dagger}(q)$, $a(q)$ and anti -particle operators  $b^{\dagger}(q)$,$ b(q)$ :
\begin{equation}
C(y)=\int_{-\infty}^{\infty}\,\frac{dq}{2\pi}\Big[a(q)|U^{+}(q)\rangle e^{-iqy}+b^{\dagger}(q)|V^{-}(q)\rangle e^{iqy}\Big]
\label{field}
\end{equation}
The Hamiltonian in the eigenvalue representation  is given by:
\begin{equation}
H^{edge}=\int_{-\infty}^{\infty}\,\frac{dq}{2\pi}\Big[a^{\dagger}(q)a(q)\Big(\hbar v| q| -E_{F}\Big)+b^{\dagger}(q)b(q)\Big(\hbar v| q| +E_{F}\Big)\Big]
\label{hamiltonian}
\end{equation}
The electric and thermal current in the static  limit $J^{el}(Q\rightarrow 0,t)$ and $J^{heat}(Q\rightarrow 0,t)$
are given by:
\begin{eqnarray}
&&J^{el}(Q\rightarrow 0,t)=ev \int_{-\infty}^{\infty} \frac{dq}{2\pi}\Big[a^{\dagger}(q)a(q)\langle U^{+}(q)|\sigma_{2}| U^{+}(q)\rangle -b^{\dagger}(q)b(q)\langle V^{-}(q)|\sigma_{2}| V^{-}(q)\rangle\Big] \nonumber\\&&
J^{heat}(Q\rightarrow 0,t)=\nonumber\\&& \hbar v^2 \int_{-\infty}^{\infty}\,\frac{dq}{2\pi}\Big[a^{\dagger}(q)a(q)\Big(q- k_{F}\langle U^{+}(q)|\sigma_{2}|U^{+}(q)\rangle \Big) -b^{\dagger}(q)b(q)\Big(q- k_{F}\langle V^{-}(q)|\sigma_{2}|V^{-}(q)\rangle\Big)\Big]
\nonumber\\&&
\end{eqnarray}
Next, we compute the matrix elements $\langle U^{+}(q)|\sigma_{2}|U^{+}(q)\rangle$ , $ \langle V^{-}(q)|\sigma_{2}|V^{-}(q)\rangle $ and introduce the energy  variables , $\hbar v|q|=\epsilon$,  $\hbar vk_{F}=E_{F}$, $\pm| q|=\frac{\pm \epsilon}{\hbar v}$.

\noindent
 For the electric currents ,we obtain  the following equations :
\begin {eqnarray}
&&J^{el}(Q\rightarrow 0,t)=\frac{e}{h}\int_0^{\infty}\,d \epsilon \Big[ \Big(a^{\dagger}(\epsilon)a(\epsilon)-a^{\dagger}(-\epsilon)a(-\epsilon)\Big)+ \Big(b^{\dagger}(\epsilon)b(\epsilon)-b^{\dagger}(-\epsilon)b(-\epsilon)\Big)\Big]\nonumber\\&&
J^{heat}(Q\rightarrow 0,t)=\nonumber\\&&\frac{1}{h}\int_0^{\infty}\,d \epsilon\Big [\Big(a^{\dagger}(\epsilon)a(\epsilon)-a^{\dagger}(-\epsilon)a(-\epsilon)\Big)\Big(\epsilon-E_{F}\Big)-\Big(b^{\dagger}(\epsilon)b(\epsilon)-b^{\dagger}(-\epsilon)b(-\epsilon)\Big)\Big(\epsilon+E_{F})\Big)\Big]\nonumber\\&&
\end{eqnarray}

\vspace{0.2in}

\textbf{IV-The electric and  thermal conductance:  a  Landauer- Buttiker approach  to the zero modes one-dimensional Dirac Fermions}

\vspace{0.2in}

We will follow the Landauer-Buttiker approach based on the $\mathbf{S}$ matrix   given by  \cite{Buttiker,Flensberg,Butcher}.  We  introduce a  modification of the  $\mathbf{S}$ matrix needed  for  Dirac materials  which contains in addition  to  particles anti-particles which are less known.  
For the present case we have no backscattering and the  transmission function is  $|t(\epsilon)|^2=1$.
We attach two reservoirs  at the boundaries $y=0 $   and $ y=L$.  $L$ obeys  $ L>>D$ and for practical reasons replace  the   limit $\int _{-L}^{L}\,dy $  with 
$\int _{-\infty}^{\infty}\,dy $. The left reservoir is  at ($y=0$) injects Fermions to the right  with momentum $q>0$. This   is represented in terms  of  the Fermi-Dirac  occupation function :
\noindent

 $\langle a^{\dagger}(\epsilon)a(\epsilon)\rangle=f^{F.D.particle}_{R}$ and $\langle  b^{\dagger}(\epsilon)b(\epsilon)\rangle =f^{F.D.anti-particle}_{R}$ . The index $R$ represents particles which move to the right with momentum  $q>0$ and energy $\epsilon$.

\noindent

 Similarly, the right reservoir (at $ y=L\rightarrow\infty$ ) injects particles with momentum $-q<0$ and corresponds to

\noindent 
 $\langle a^{\dagger}(-\epsilon)a(-\epsilon)\rangle=f^{F.D.particle}_{L}$ and
$\langle b^{\dagger}(-\epsilon)b(-\epsilon)\rangle=f^{F.D.anti-particle}_{L}$.

\noindent
The index $L$ represents the  mover to the left with momentum $-q<0$ and energy $-\epsilon$.

\noindent 
We will  first consider  the $ \mathbf{electrical}$ conductance.
 
\noindent
We attach the left reservoir to a  source with  a voltage $V_{G}$ and the right reservoir to a source with zero voltage  $V_{G}=0$. At low temperatures we find:
\begin{eqnarray}
&& \langle a^{\dagger}(\epsilon)a(\epsilon)\rangle=f^{F.D.particle}_{R}=f^{F.D.}(\epsilon -E_{F}-eV_{g}), \hspace{0.1 in} \langle a^{\dagger}(-\epsilon)a(-\epsilon)=f^{F.D.particle}_{L}=f^{F.D.}(\epsilon -E_{F})\nonumber\\&&
\langle b^{\dagger}(\epsilon)b(\epsilon)\rangle=f^{F.D.anti-particle}_{R}=f^{F.D.}(\epsilon +E_{F}+eV_{g}), \hspace{0.1 in} \langle b^{\dagger}(-\epsilon)b(-\epsilon)=f^{F.D.anti-particle}_{L}=f^{F.D.}(\epsilon +E_{F})\nonumber\\&&
\end{eqnarray}
The electric current is given by:
\begin{eqnarray}
&&\langle J^{el}(Q\rightarrow 0 )\rangle\nonumber\\&&=\frac{e}{h}\int_{0}^{\infty}d\epsilon\Big[\Big(\langle a^{\dagger}(\epsilon)a(\epsilon)\rangle-\langle a^{\dagger}(-\epsilon)a(-\epsilon)\Big)+\Big(\langle b^{\dagger}(\epsilon)b(\epsilon)\rangle-\langle b^{\dagger}(-\epsilon)b(-\epsilon)\Big)\Big]\nonumber\\&&=\frac{e}{h}\int_{0}^{\infty}d\epsilon\Big[\Big(f^{F.D.}(\epsilon -E_{F}-eV_{g})-f^{F.D.}(\epsilon -E_{F})\Big)+\Big(f^{F.D.}(\epsilon+ E_{F}+eV_{g})-f^{F.D.}(\epsilon+ E_{F})\Big)\Big]\nonumber\\&& =\frac{e^2}{h}V_{g}\int_{0}^{\infty}d\epsilon \Big(\delta (\epsilon-E_{F})-\delta( \epsilon+E_{F})\Big)=\frac{e^2}{h}V_{g}\Big(\theta[E_{F}]-\theta[-E_{F}]\Big)\nonumber\\&&
\end{eqnarray}

\noindent
 The electrical conductance  is    $G=\frac{e^2}{h}$  for positive Fermi energy (electron -like) and negative for the   negative Fermi energy (hole-like).

\noindent
Next, we consider the $\mathbf{thermoelectric}$ conductance .

\noindent
 We attach  a  thermal reservoir  at  temperature $T+\Delta T$ to the left   and a    reservoir at   temperature  $T$ to the right  \cite{Mahan,Butcher,Flensberg}.
Considering the particle and  anti-particle contributions,  we find the following  equation for the thermal conductance:
\begin{eqnarray}
&&  \langle a^{\dagger}(\epsilon)a(\epsilon)\rangle=f^{F.D.particle}_{R}=f^{F.D.}\Big(\frac{\epsilon -E_{F}}{K_{B}(T+\Delta T)}\Big), \hspace{0.1 in} \langle a^{\dagger}(-\epsilon)a(-\epsilon)\rangle=f^{F.D.particle}_{L}=\Big(\frac{\epsilon -E_{F}}{K_{B}T}\Big)\nonumber\\&&
\langle b^{\dagger}(\epsilon)b(\epsilon)\rangle=f^{F.D.anti-particle}_{R}=f^{F.D.}\Big(\frac{\epsilon+ E_{F}}{K_{B}(T+\Delta T)}\Big),\nonumber\\&&\langle b^{\dagger}(-\epsilon)b(-\epsilon)=f^{F.D.anti-particle}_{L}=f^{F.D.}\Big(\frac{\epsilon+ E_{F}}{K_{B}T}\Big)\nonumber\\&&
\end{eqnarray}
The electric current at finite temperatures is:
\begin{eqnarray}
  &&\langle J^{el}(Q\rightarrow 0 )\rangle=\nonumber\\&&\frac{e}{h}\int_{0}^{\infty}d\epsilon\Big[\Big(\langle a^{\dagger}(\epsilon)a(\epsilon)\rangle-\langle a^{\dagger}(-\epsilon)a(-\epsilon)\Big)+\Big(\langle b^{\dagger}(\epsilon)b(\epsilon)\rangle-\langle b^{\dagger}(-\epsilon)b(-\epsilon)\Big)\Big]\nonumber\\&&=\frac{e}{h}\int_{0}^{\infty}d\epsilon\Big[ f^{F.D.}\Big(\frac{\epsilon -E_{F}}{K_{B}(T+\Delta T)}\Big)-f^{F.D.}\Big(\frac{\epsilon -E_{F}}{K_{B}T}\Big)+f^{F.D.}\Big(\frac{\epsilon+ E_{F}}{K_{B}(T+\Delta T)}\Big)-f^{F.D.}\Big(\frac{\epsilon+ E_{F}}{K_{B}T}\Big)\Big]\nonumber\\&&
=\frac{eK_{B}}{h}(-\Delta T) \cdot\Big[ \int_{\frac{-E_{F}}{K_{B}T}}^{\infty} x\partial_{x} f^{F.D.}(x) dx +  \int_{\frac{E_{F}}{K_{B}T}}^{\infty} x\partial_{x} f^{F.D.}(x) dx \Big]\nonumber\\&&
\end{eqnarray}
The $\mathbf{thermoelectric}$ current is given in Fig.1. The figure shows that  the thermoelectric  conductance vanishes for temperature  and  Fermi energy which obey $\frac{E_{F}}{K_{B}T}>7$.
The $\mathbf{thermal}$ conductance is:
\begin{eqnarray}
&& \langle  J^{heat}(Q\rightarrow 0 )\rangle= \nonumber\\&&
\frac{1}{h}\int_{0}^{\infty}d\epsilon \Big[\Big(\langle a^{\dagger}(\epsilon)a(\epsilon)\rangle-\langle a^{\dagger}(-\epsilon)a(-\epsilon)\Big)\Big(\epsilon-E_{F}\Big)-\Big(\langle b^{\dagger}(\epsilon)b(\epsilon)\rangle-\langle b^{\dagger}(-\epsilon)b(-\epsilon)\Big)\Big(\epsilon+E_{F}\Big)\Big]
\nonumber\\&&
\approx  \frac{ K^{2}_{B} T}{h}(-\Delta T)\cdot\Big[\int_{\frac{-E_{F}}{K_{B}T}}^{\infty}  x^2 \partial_{x} f^{F.D.}(x) dx+ \int_{\frac{E_{F}}{K_{B}T}}^{\infty} x^2 \partial_{x} f^{F.D.}(x) dx\Big]\nonumber\\&&
\end{eqnarray}
In Fig.2, we show the $\mathbf{thermal}$ conductance.
The thermal conductance is half of the value of the  free electrons.
 In this case ,  the conductance $G$, the  thermal conductance $kappa$ and  the thermoelectric conductance  $L$  are half the value obtained for regular electrons\cite{Kamran} . We find the  values $G=\pm\frac{e^2}{h}$, $\kappa=\frac{ \pi^2}{3h}K^2_{B}T$ and $L=\frac{e}{h}K_{B}\cdot I[\frac{E_{F}}{K_{B}T}] $  with   $I[\frac{E_{F}}{K_{B}T}]$ decreasing from one to zero for $\frac{E_{F}}{K_{B}T}>7$.

\vspace{0.2in}

\textbf{IVb- The thermoelectric effects in the presence of a magnetic field $B$ and  backscattering}

\vspace{0.2in}

In this section we will consider the two dimensional model in the presence of a magnetic field with backscattering .This investigation will be based on the one dimensional model given in  Eq.$6$  obtained from the projection introduce in section II.  
In the presence of a magnetic field, backscattering is allowed and  will contribute to the electric and thermal   conductance.
We include in  the previous model the magmetic field \textbf{$ B$  in the $x$ direction} and a potential $V(y)\approx V^{0}(y)+V^{back}(y)$. $ V^{0}(y)$ represents the forward scattering and $V^{back}(y)$ is the backscattering potential.
The  Hamiltonian is given by: 
\begin{eqnarray}
&& H^{edge}= H+H^{disorder}\nonumber\\&&
H=\hbar v \int_{-\infty}^{\infty} dy \Big[C^{\dagger}(y)\Big( \sigma_{2}(-i\partial_{2}) +\sigma_{1}B-I k_{F}\Big)C(y)\Big]\nonumber\\&&
H^{disorder}= \int_{-\infty}^{\infty}dy \Big[C^{\dagger}(y)\Big(I\cdot V^{0}(y)+ I\cdot V^{back}(y)\Big)C(y)\Big]\nonumber\\&&
\end{eqnarray}
We find the eigen spinor $|\hat {U}^{+}(\epsilon)\rangle=\frac{1}{\sqrt{2}}\Big[e^{-i\varphi},1\Big]^{T}$
and $|\hat {V}^{-}(\epsilon)\rangle=\frac{1}{\sqrt{2}}\Big[e^{i\varphi},1\Big]^{T}$, where $\tan[\varphi]=\frac{\epsilon}{B}$. The spinor allows for backscattering .We find $\langle\hat {U}^{+}(\epsilon)| \hat {U^{+}}(\epsilon)\rangle=\langle\hat {V}^{-}(\epsilon)| \hat {V}^{-}(\epsilon)\rangle= \frac{e^{i2\varphi}+1}{2}$
The Hamiltonian $H$  given in equation $19$  becomes the following  in the eigenvalue form  :
\begin{eqnarray}
&&H=\int_{0}^{\infty}d\epsilon\Big[\Big(a^{\dagger}(\epsilon)a(\epsilon)+a^{\dagger}(-\epsilon)a(-\epsilon)\Big)\Big(E-E_{F}\Big)+\Big(b^{\dagger}(\epsilon)b(\epsilon)+b^{\dagger}(-\epsilon)b(-\epsilon)\Big)\Big(E+E_{F}\Big)\Big]\nonumber\\&&
E=\sqrt{\epsilon^{2}+B^2}\nonumber\\&&
\end{eqnarray}
The  electric current operator for the Hamiltonian given in Eq.$(19)$ takes the form:
\begin{eqnarray} 
&&  J^{el}(Q\rightarrow 0 )=ev\int_{-\infty}^{\infty}\frac{dq}{2\pi}\Big[a^{\dagger}(q)a(q)\langle \hat{ U^{+}}(q)|\sigma_{2}|\hat{ U}^{+}(q)\rangle+b^{\dagger}(q)b(q)\langle \hat{V}^{-}(q)|\sigma_{2}|\hat{ V}^{-}(q)\rangle\Big]\nonumber\\&&
\end{eqnarray}
The effect of  the forward scattering  $ V^{0}(y) $ is taken  into account by introducing the   inverse life  time $\eta$ and replacing  the energy with  $E\rightarrow E \pm i\eta$. The effect of the backscattering is included in  the  construction of  the reflected states. An incoming state $|q\rangle$ is reflected back with the amplitude $r(-q;q)$ . The reflected state is   $r(-q;q)|-q\rangle$.
To compute this amplitude, we use  the equation of motions for the backscattering potential:
\begin{eqnarray}
&&\partial_{t}a(q,t)=\frac{1}{i\hbar}a(q,t)\Big(E-E_{F}+i\eta\Big)+\frac{1}{i\hbar}V^{back}(2q)\hat{ U}^{+}(q)||\hat{ U}^{+}(-q)\rangle a(-q,t)\nonumber\\&&
a(q,t)\approx a_{0}(q,t)+\delta a(q,t);\hspace{0.1in}\partial_{t}a_{0}(q,t)=\frac{1}{i\hbar}a_{0}(q,t)\Big(E-E_{F}+i\eta\Big),\nonumber\\&&\partial_{t}\delta a(q,t)=\frac{-i}{\hbar}\delta(q,t)\Big(E-E_{F}+i\eta\Big)-i\frac{V^{back}(2q)}{\hbar}\langle\hat{ U}^{+}(q)|\hat{ U}^{+}(-q)\rangle  a(-q,t)\nonumber\\&&
\end{eqnarray}
We will solve the equation using the Fourier transform $\delta a(q,t)=\int\frac{d\omega}{2\pi}\delta a(q,\omega)e^{-i\omega t}$. This gives the equation :$\delta a(q,\omega)=\frac{1}{\hbar}\frac{V^{back}(2q)\langle\hat{ U}^{+}(q)|\hat{ U}^{+}(-q)\rangle}{\omega -((E-E_{F} )\pm i \eta)} a(-q,\omega)$.
As a result, the transmission intensity is given in terms of the matrix elements $\langle\hat{ U}^{+}(q)|\hat{ U}^{+}(-q)\rangle$ for particles and $\langle\hat{ V}^{-}(q)|\hat{ V}^{-}(-q)\rangle$ for anti-particles by :
\begin{eqnarray}
&&t^{2}(E;\omega,+)=1-|r(-q;q|\omega)|^2=1-\frac{1}{\hbar^2}\frac{|V^{back}(2q)|^2|\langle\hat{ U}^{+}(q)|\hat{ U}^{+}(-q)\rangle|^2}{(\omega -(E-E_{F})) ^2+\eta^2)}\nonumber\\&&
t^{2}(E;\omega,-)=1-|r(-q;q|\omega)|^2=1-\frac{1}{\hbar^2}\frac{|V^{back}(2q)|^2|\langle\hat{ V}^{-}(q)|\hat{ V}^{-}(-q)\rangle|^2}{(\omega -(E+E_{F})) ^2+\eta^2)}\nonumber\\&&
\end{eqnarray}
The Hamiltonian  for the  two momentum $ |q\rangle$ and $|-q\rangle$ with the non -perturbed  energy $ E(q)=E(-q)=E$ is given by  :
\begin{equation}
 E(q)\Big(|q\rangle   \langle q|+ |-q\rangle \langle -q| \Big) + V^{back}(2q)\langle\hat{ U^{+}}(q)|\hat{ U^{+}}(-q)\rangle|q\rangle  \langle-q | +(V^{back}(2q))^{*}\langle\hat{ U^{+}}(-q)|\hat{ U^{+}}(q)\rangle |-q\rangle   \langle  q|
\label{equation} 
\end{equation} 
The lowest eigenvalue of this Hamiltonian  is given by:
 $\omega=E-|V^{back}(2q)| \cos^2{\varphi(E)}$. We substitute in the transmission function    $ \omega=(E-E_{F})-|V^{back}(2q)| \cos^2{\varphi(E)}$ for particles and  $ \omega=(E+E_{F})-|V^{back}(2q)| \cos^2{\varphi(E)}$ for anti-particles. We obtain:
\begin{equation}
t^{2}(E;\omega,+)=t^{2}(E;\omega,-)=t^2(E)=1-\frac{\cos^2{\varphi(E)}}{\cos^4{\varphi(E)}+(\frac{\eta}{|V^{back}(2q)|})^2}
\label {tansmission}
\end{equation}
The electric current operator given in Eq.$ 21$ takes the form:
\begin{eqnarray}
&&\langle J^{el}(Q\rightarrow 0 )\rangle \nonumber\\&&=\frac{e}{h}\int_{0}^{\infty}dE(\frac{E}{\epsilon}) \sin{\varphi(E)}t^2(E)\cdot\Big[\Big(\langle a^{\dagger}(E)a(E)\rangle-\langle a^{\dagger}(-E)a(-E)\Big)+\Big(\langle b^{\dagger}(E)b(E)\rangle-\langle b^{\dagger}(-E)b(-E)\Big)\Big]\nonumber\\&&
t^2(E)=1-\frac{\cos^2{\varphi(E)}}{\cos^4{\varphi(E)}+(\frac{\eta}{|V^{back}(2q)|})^2}=1-\frac{(\frac{B}{E})^2}{(\frac{B}{E})^4+(\frac{\eta}{|V^{back}(2q)|})^2}\nonumber\\&& 
\end{eqnarray}
\vspace{0.1 in}
The change of the integration measure from $d\epsilon$  to $dE$ gives rise  to the product  $(\frac{E}{\epsilon}) \cdot \sin{\varphi(E)}$ which is equal to 1.

\noindent
The electrical current due to a voltage $eV_{g}$ at $ T\rightarrow 0$ is given  in terms of $t^2(E)$ :
\begin{eqnarray}
&&\langle J^{el}(Q\rightarrow 0 )\rangle=\nonumber\\&&\frac{e}{h}\int_{0}^{\infty}d E t^2(E)\Big[\Big(f^{F.D.}E -E_{F}-eV_{g})-f^{F.D.}(E -E_{F})\Big)+\Big(f^{F.D.}(E+ E_{F}+eV_{g})-f^{F.D.}(E+ E_{F})\Big)\Big]\nonumber\\&& =\frac{e^2}{h}V_{g}\int_{0}^{\infty}dE t^2(E) \Big (\delta (E-E_{F})-\delta( E+E_{F})\Big)=\frac{e^2}{h}V_{g}\Big(t^2(E_{F})\theta[E_{F}]-\theta[-E_{F}] t^2(-E_{F})\Big)\nonumber\\&&
\end{eqnarray}
This result is shown in Fig. 3,where  the magnetic field and   backscattering potential decrease the conductance.  We see that the  $\mathbf{ electric}$ conductance $ G[\frac{B}{E_{F}}]$ decreases  with the increase  in  the ratio  $\frac{B}{E_{F} }$ .

\noindent
The $\mathbf{ thermoelectric}$ conductance is given in terms of the transmission function $ t^{2}(E)$:
\begin{eqnarray}
&&\langle J^{el}(Q\rightarrow 0 )\rangle=\nonumber\\&&\frac{e}{h}\int_{0}^{\infty}d E  t^{2}(E)\Big[ f^{F.D.}\Big(\frac{E -E_{F}}{K_{B}(T+\Delta T)}\Big)-f^{F.D.}\Big(\frac{E -E_{F}}{K_{B}T}\Big)+f^{F.D.}\Big(\frac{E+ E_{F}}{K_{B}(T+\Delta T)}\Big)-f^{F.D.}\Big(\frac{E+ E_{F}}{K_{B}T}\Big)\Big]\nonumber\\&&
\end{eqnarray}
The thermoelectric current decreases with the increase in the magnetic field. This is seen in Fig.4. where the upper graph shows  the plot for $B=0$  and the lower graph represents the thermoelectric current for $\frac{B}{E_{F}}=1$. 
 
\noindent
The $\mathbf{ thermal}$ current  is  obtained from the continuity equation. We obtain the $\mathbf{ thermal}$ current in the presence of the magnetic field $B$ in the $x$ direction :

\noindent
$J^{heat}(y,t)=\hbar v^2 C^{\dagger}(y)\Big(-i\partial_{2} -k_{F}\sigma_{2}  -iB\sigma_{1}\Big)C(y)$

\noindent
This equation is used   together with the  transmission function $t^2(E)$=$1-\frac{(\frac{B}{E})^2}{(\frac{B}{E})^4+(\frac{\eta}{|V^{back}(2q)|})^2}$.  We observe that the spinors $ |\hat {U}^{+}(\epsilon)\rangle \frac{1}{\sqrt{2}}\Big[e^{i\varphi},1\Big]^{T}$, $ |\hat {V}^{-}(\epsilon)\rangle\frac{1}{\sqrt{2}}\Big[e^{-i\varphi},1\Big]^{T}$ give $ \langle \sigma_{1}\rangle =0 $ and  $\langle\hat {U}^{+}(\epsilon)|\sigma_{2} |\hat {U}^{+}(\epsilon)\rangle =\sin[\varphi(E)]$.
As a result, the thermal current is given by:
\begin{eqnarray}
&& \langle  J^{heat}(Q\rightarrow 0 )\rangle= \nonumber\\&&\frac{1}{h}\int_{0}^{\infty}d E \frac{E}{\sqrt {E^2-B^2}} t^{2}(E)\Big[ f^{F.D.}\Big(\frac{E -E_{F} \cdot\sin[\varphi(E)]}{K_{B}(T+\Delta T)}\Big)-f^{F.D.}\Big(\frac{E -E_{F} \cdot\sin[\varphi(E)]}{K_{B}T}\Big)\nonumber\\&&+f^{F.D.}\Big(\frac{E+ E_{F} \cdot\sin[\varphi(E)]}{K_{B}(T+\Delta T)}\Big)-f^{F.D.}\Big(\frac{E+ E_{F} \cdot\sin[\varphi(E)]}{K_{B}T}\Big)\Big]\nonumber\\&&
\end{eqnarray}
In Fig.5. we show the thermal conductance as a function of the magnetic field $B$.  it is found  that when  the magnetic field increases, the thermal conductance decreases.

\noindent 
Notably, for the $3D$ topological  insulator with a $2D$ boundary, the measurements at low temperature resistance  \cite{Ando,Kozlov}  show an increase of the  resistance  with the increases in the magnetic field.
Applying our theory to this case, we show  an  increase in  the resistance in Fig.6 with the magnetic field.Our theory is formulated  at  finite temperatures  where weak anti-localization effects can be ignored   \cite{Raman}. As a result, backscattering controls the conductance.
In $Appendix$ $A $ we present the derivation of the backscattering inverse  life  time for the $2D$ boundary.

\vspace{0.2in}

\textbf{VI- Proposed experimental set up for testing thermoelectricity }

\vspace{0.2in}

We consider a two dimensional time reversal invariant topological insulator in HgTe/CdTe quantum wells. We  obtain topologically protected edge states . We consider a sample with the width in the $x$ direction   larger than  $D>1000nm$ \cite{Zhou}. Under these  conditions,  the edge mode at  $x=0$  is not  affected by other  boundaries. The  length in the $y$ direction $ L$ is much larger than the width $D$ in the $x$ direction .According to the experimental observation for  HgTe Quantum wells the inter-edge tunneling is negligible.  A temperature gradient or voltage difference $eV_{g}$ along the $ y$  direction is achived   in the following way.  On the left side of the sample, we apply a heating device which will create a temperature $T+\Delta T $ or apply a voltage  $V_{g}$ .The right side of the sample is   kept at temperature $T$ or voltage $ V_{g}=0$.
As a result,   we will observe a thermoelectric voltage between the left and right sides  of the sample, or a current driven by $ V_{g}$.
Due to backscattering in the  magnetic field the thermoelectric and electrical  resistance  increase with the increase of the magnetic field  (see Figures $3-5 $).

\vspace{0.2 in}

\textbf{VII-Conclusion}

\vspace{0.2 in}

The  thermoelectricity for the  zero modes was  computed. The heat current and the electric current are obtained  from the Hamiltonian  and  Heisenberg equation of motion. The edge mode of a $2D $ and $3D$  topological model is investigated. For $D$ we obtain an effective one dimensional model  and for $D=3$ we have an effective two dimensional model. In the the presence of a magnetic field, the backscattering is enhanced.  We find that the  electric, thermoelectric  and  thermal resistance increase with the increases of the magnetic field. For $3D$ topological insulator a finite temperature we can ignore weak localization effects. Using  only backscattering , we confirm the experimental result obtained for the resistance in magnetic fields  \cite{Ando,Kozlov}.An  experimental set up  was proposed to test our theory.

\vspace{0.2 in}

$Appendix -A$

The $ 2D$ boundary surface with a magnetic field $ B$ in the $z$ direction is given by the Hamiltonian  :

\vspace{0.2 in}

\begin{equation}
H^{3D}\Rightarrow H^{2Dedge}= \hbar v \int\,d^2r \Big[C^{\dagger}(\vec{r})\Big( \sigma_{1}(-i\partial_{1}) +\sigma_{2}(-i\partial_{2})-\sigma_{3}B -I k_{F}\Big)C(\vec{r})\Big] 
\label{equation}
\end{equation}

Using this Hamiltonian, we compute the backscattering   lifetime. The assumption in this calculation is that   it is possible to extract the backscattering potential $ V^{back}(\vec{K}_{F};-\vec{K}_{F} +\vec{q})$ where $|\vec{q}|<|\vec{Q}|<2K_{F}$. Due to lattice effects the density state is maximal  and   $ 2K_{F}$ scattering    dominantes: 
\begin{eqnarray}
&&\frac{1}{\tau^{back}_{K}}= 2 \pi\int_{\vec{K}_{F}}^{-\vec{K}_{F}+\Delta \vec{Q}}\frac{d^2q}{(2\pi)^2}|V^{back}\langle \hat  {U}^{+}(\vec{K}_{F)}|\hat{ U}^{+}(-\vec{K}_{F}+\vec{q}\rangle|^2\delta \Big(E(\vec{K}_{F})-E(-\vec{K}_{F}+\vec{q})\Big)\nonumber\\&&\approx \Big(1 - \frac{1}{1 + (\frac{B}{E_{F}})^2 - \frac{B}{E_{F}}\sqrt{1+ \frac{B}{E_{F}}}}\Big)^2\nonumber\\&&
\end{eqnarray}
The magnetic field is measured in units of energy such that $\frac{B}{E_{F}}$  is dimensonless

\clearpage

\begin{figure}
\begin{center}
\includegraphics[width=4.5 in]{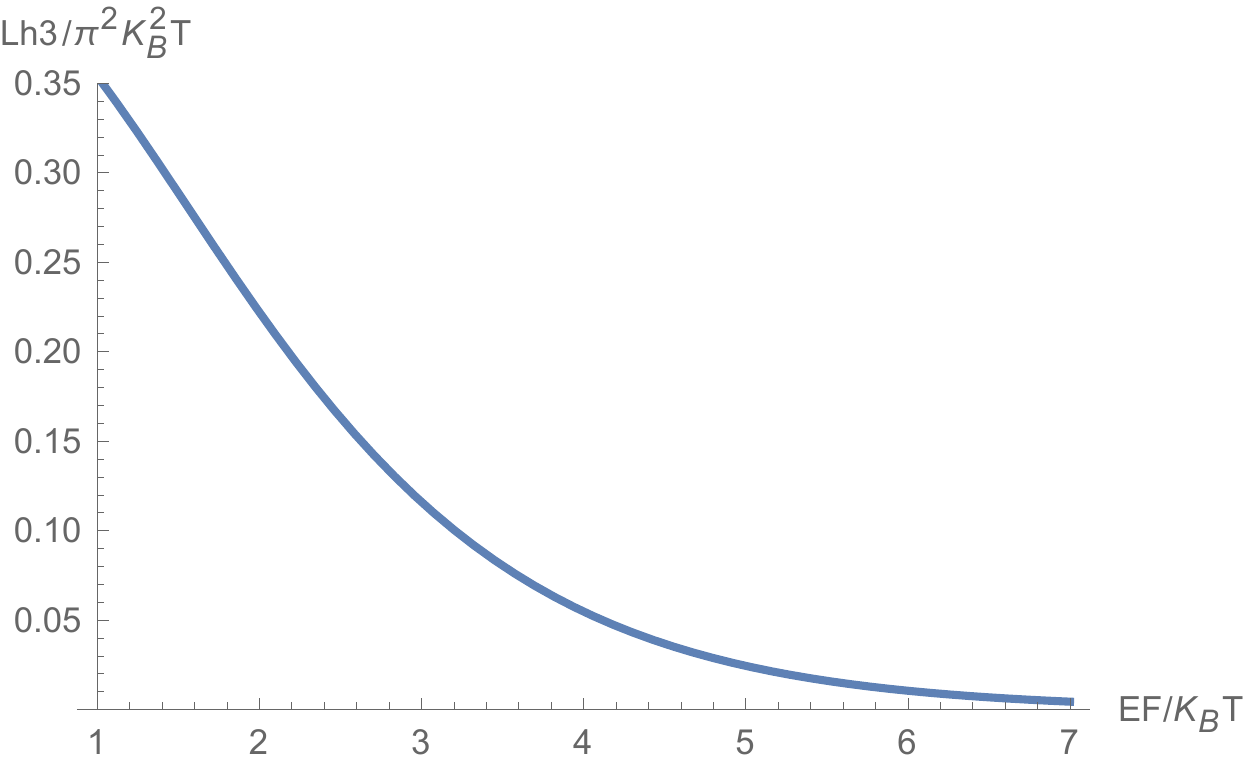}
\end{center}
\caption{The thermoelectric current as a function of $\frac{EF}{K_{B}T}$     }
\end{figure}

\begin{figure}
\begin{center}
\includegraphics[width=4.5 in]{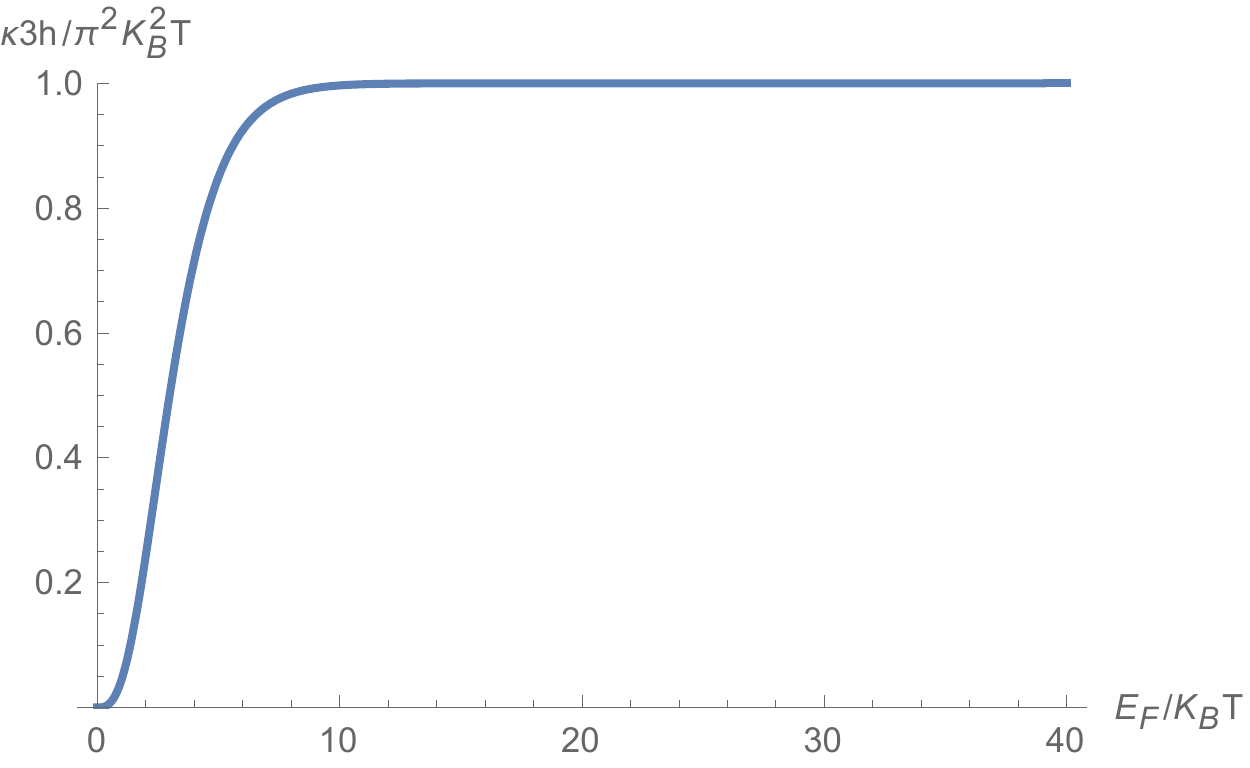}
\end{center}
\caption{The thermal   conductance      $\kappa\frac{\pi^2K^{2}_{B}T}{3h}$ as a function of $\frac{EF}{K_{B}T}$ }
\end{figure}

\begin{figure}
\begin{center}
\includegraphics[width=4.5 in]{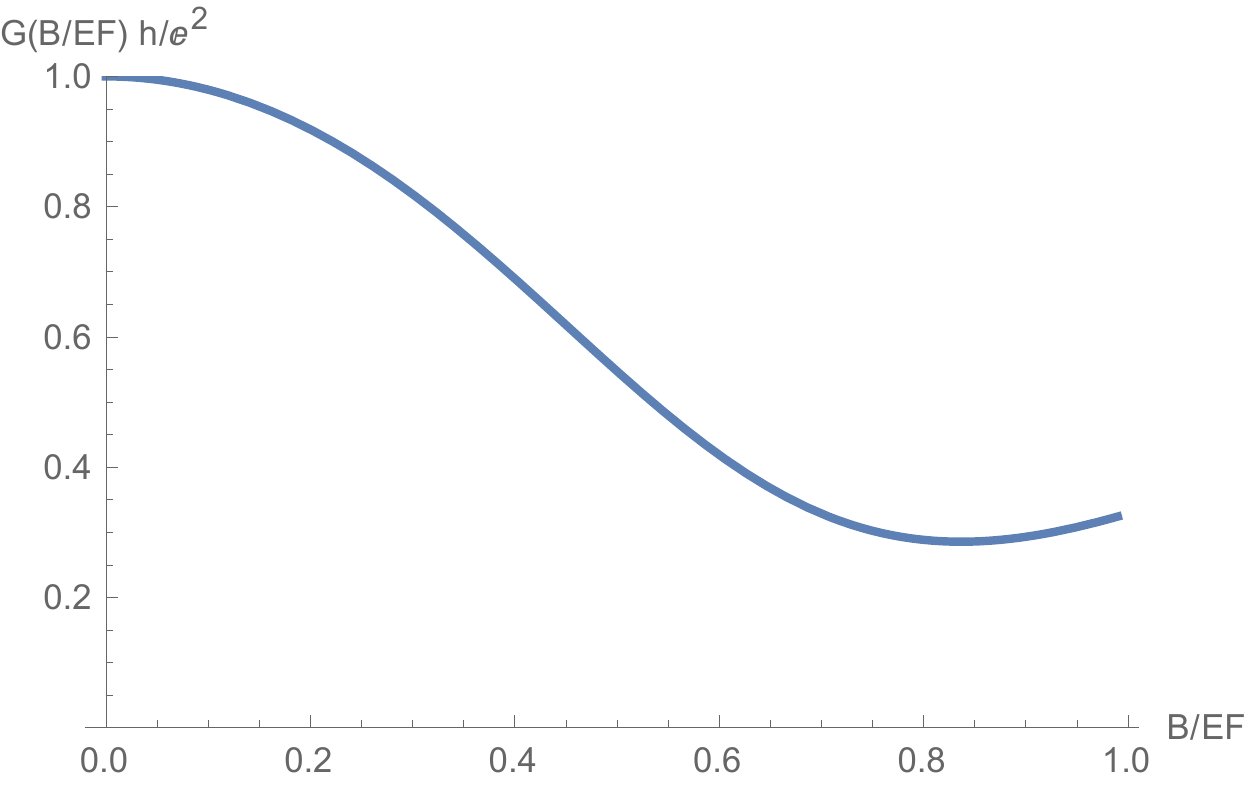}
\end{center}
\caption{The conductance of the zero mode  $G[B/EF]$  as a function of the magnetic field and Fermi energy $E_{F}$ ,$ \frac{B}{E_{F}}$  for the $2D$ topological insulator}
\end{figure}

\begin{figure}
\begin{center}
\includegraphics[width=4.5 in]{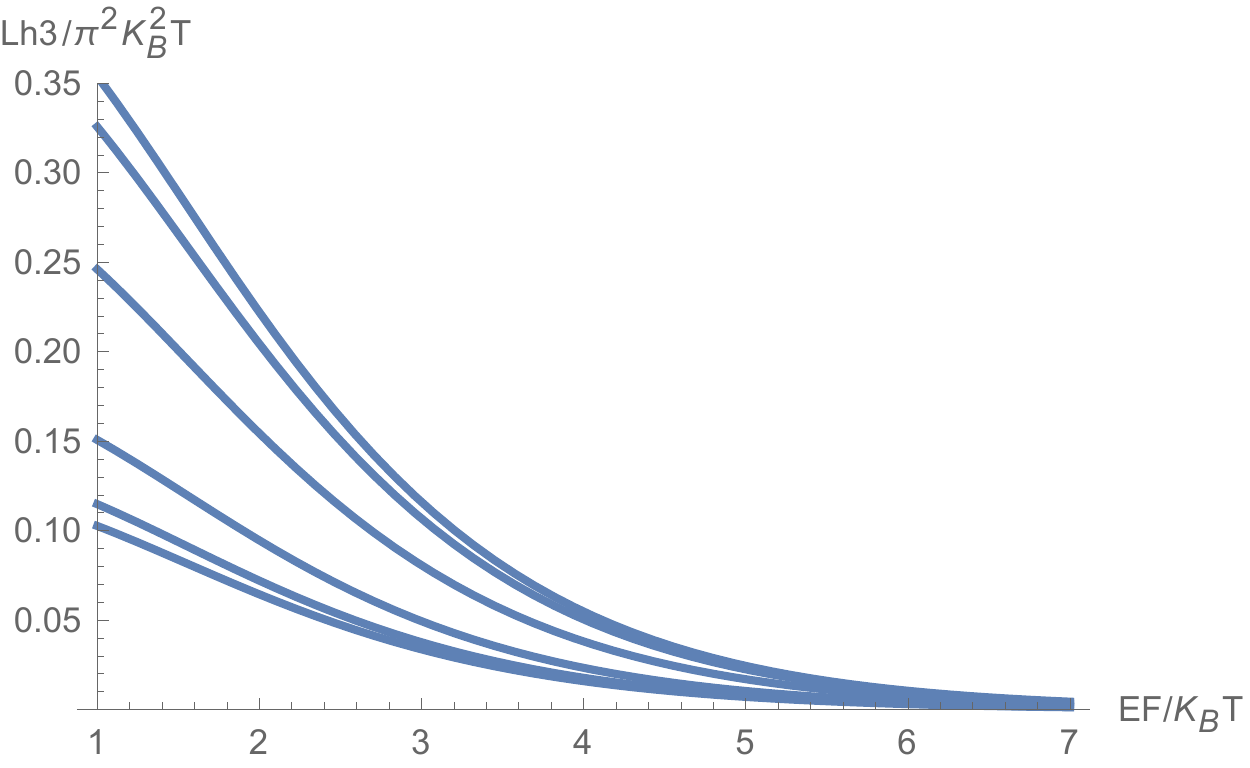}
\end{center}
\caption{The thermoelectric current as a function $\frac{EF}{K_{B}T}$  for different values of $ \frac{B}{E_{F}}$. The uper plot is for$ B=0$ and the lower plot is for $\frac{B}{E_{F}}=1$.   }
\end{figure}

\begin{figure}
\begin{center}
\includegraphics[width=4.5 in]{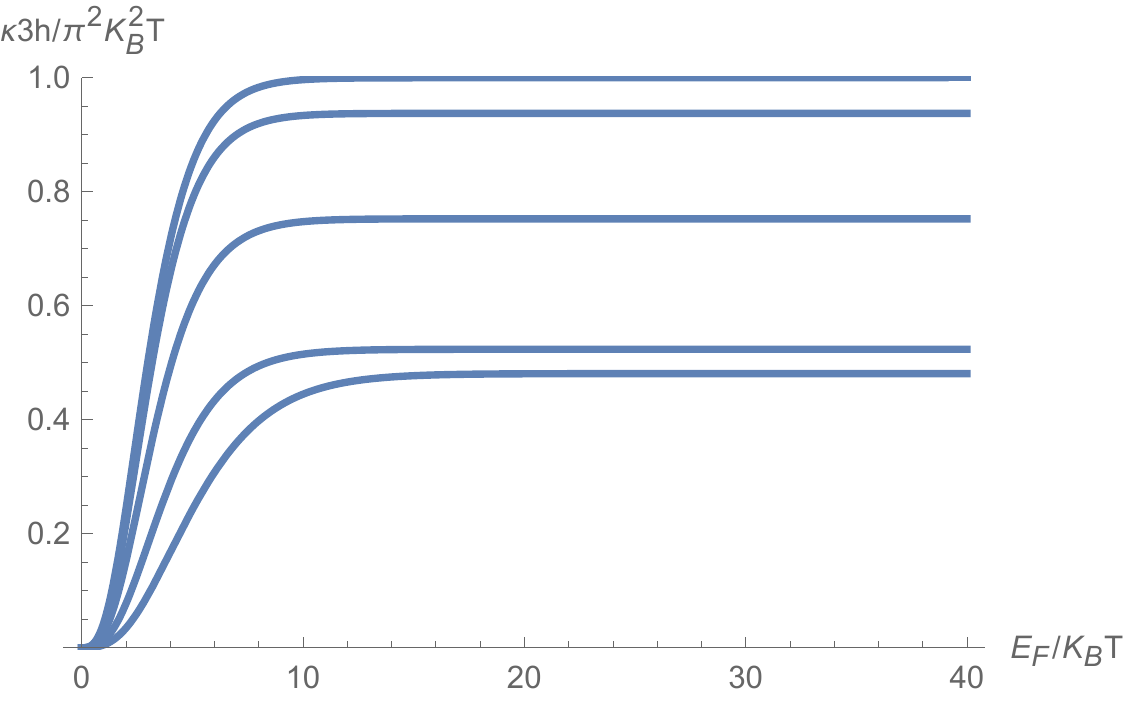}
\end{center}
\caption{The thermal   conductance      $\kappa(\frac{B}{E_{F}})\frac{\pi^2K^{2}_{B}T}{3h}$ as function  $\frac{EF}{K_{B}T}$ for increasing   values of  $ \frac{B}{E_{F}}$, the upper  plot corresponds to zero magnetic field and the lower plot is for $\frac{B}{E_{F}}=1$. }
\end{figure}

\begin{figure}
\begin{center}
\includegraphics[width=4.5 in]{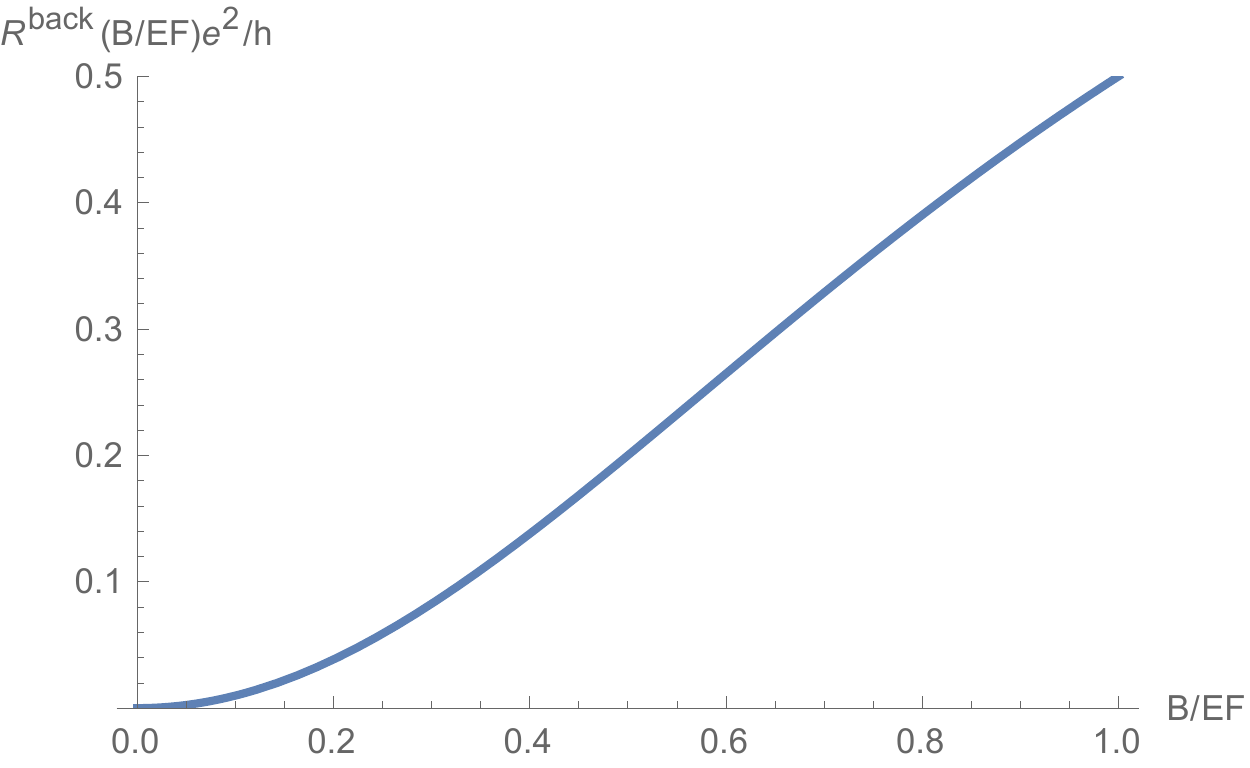}
\end{center}
\caption{The electrical    conductance   for the $3D$ topological insulator  with a $2D$ boundary (see the \cite{Kozlov} experiment)   as function  of  $\frac{EF}{K_{B}T}$ for increasing   values of  $ \frac{B}{E_{F}}$. }
\end{figure}

\begin{figure}
\begin{center}
\includegraphics[width=4.5 in]{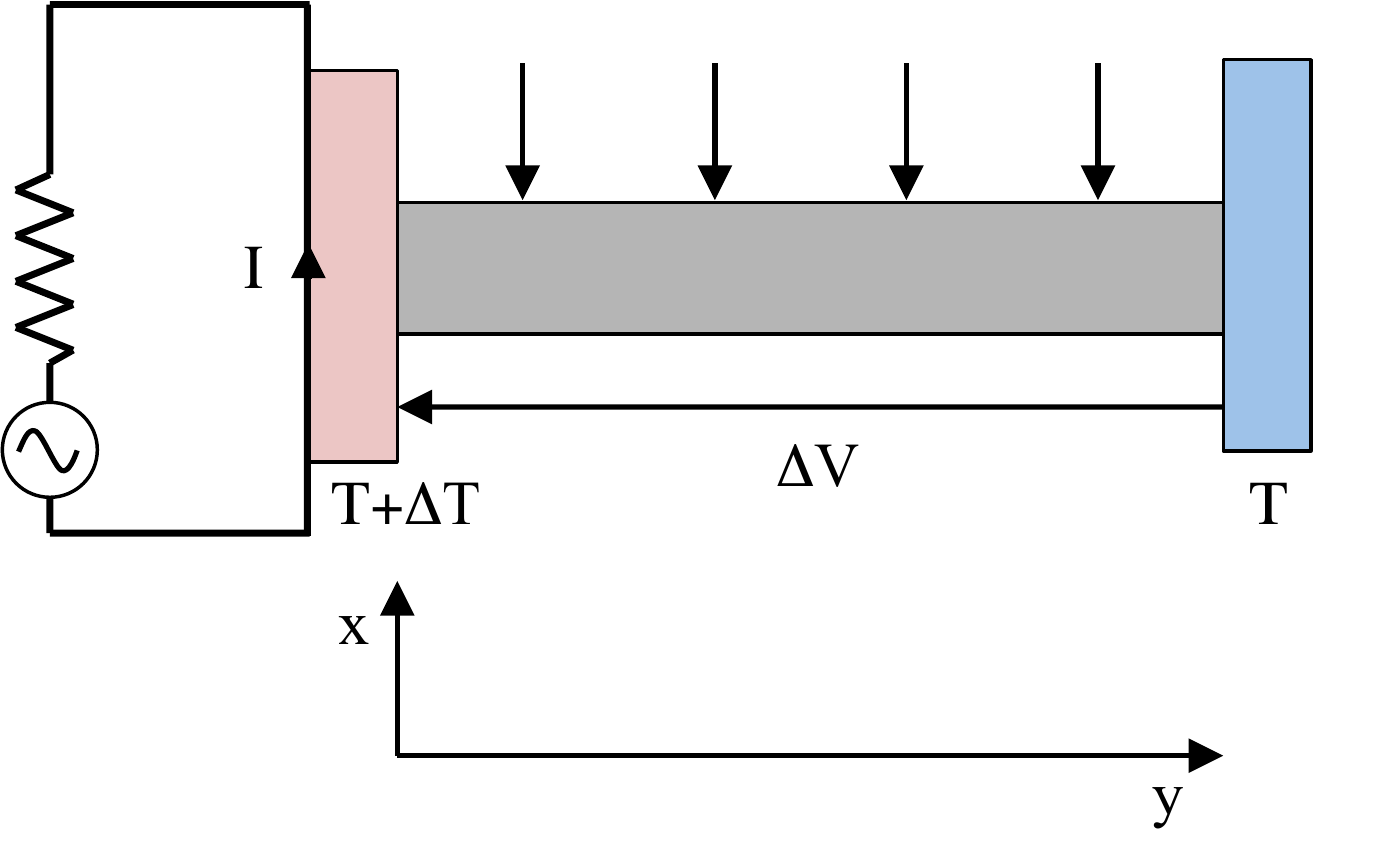}
\end{center}
\caption{The proposed experiment. The  effect of the magnetic field  is shown by the arrow in the $x$ direction. The  presence of the impurity gives rise to backscattering.   The left side of the sample is connected to a current $ I$ to  create an elevated  temperature $T+\Delta T$ with respect to the right side. A voltage $ \Delta V$ is induced by the temperature difference according to the results predicted in Fig.4. When we replace the heating current with a voltage reservoir we can measure the electrical conductance as function of magnetic field as shown in Fig.3. }
\end{figure}
\end{document}